%% file: paper.tex
\documentclass[referee]{raa}            

\usepackage{graphicx,times}             
\usepackage{natbib}
\usepackage{amssymb,amsmath}
\bibpunct{(}{)}{;}{a}{}{,}
\include{newcommands}

\usepackage[pagebackref=true,driverfallback=dvipdfm]{hyperref}
\DeclareMathOperator\erf{erf}
\hypersetup{colorlinks = true, linkcolor = green, anchorcolor = red, citecolor = blue, filecolor = red, urlcolor = red}
\usepackage{color}
\def\blue{\textcolor{black}}

\begin{document}

   \title{Subcritical dynamo and hysteresis in a Babcock-Leighton type kinematic dynamo model

}

   \volnopage{Vol.0 (20xx) No.0, 000--000}      
   \setcounter{page}{1}          

   \author{Vindya Vashishth
      \inst{1}
    \and Bidya Binay Karak
      \inst{1}
    \and Leonid Kitchatinov
      \inst{2,3}
   }

   \institute{$^1$Department of Physics, Indian Institute of Technology (BHU),
             Varanasi 221005, India; \\
             $^2$Institute of Solar-Terrestrial Physics SB RAS, Lermontov Str. 126A, 664033, Irkutsk, Russia; \\
             $^3$Pulkovo Astronomical Observatory, Pulkovskoe Sh.65, St. Petersberg, 196140, Russia
             {\it vindyavashishth.rs.phy19@itbhu.ac.in}\\
\vs\no
   {\small Received~~20xx month day; accepted~~20xx~~month day}}

\abstract{
In Sun and sun-like stars, it is believed that the cycles of the large-scale magnetic field are produced due to the existence of differential rotation and helicity in the plasma flows in their convection zones (CZs). Hence, it is expected that for each star, there is a critical dynamo number for the operation
of a large-scale dynamo. As a star slows down, it is expected that the large-scale dynamo ceases to operate above a critical
rotation period. In our study, we explore the possibility of the operation of the dynamo in the subcritical region using the
Babcock--Leighton type kinematic dynamo model. In some parameter regimes, we find that the dynamo shows hysteresis
behavior, i.e., two dynamo solutions are possible depending on the initial parameters---decaying solution if started with weak
field and strong oscillatory solution (subcritical dynamo) when started with a strong field. However, under large fluctuations in the dynamo parameter, the subcritical dynamo mode is unstable in some parameter regimes. Therefore, our study supports the possible existence of subcritical dynamo in
some stars which was previously shown in a mean-field dynamo model with distributed $\alpha$ and MHD turbulent dynamo simulations.
}

   \authorrunning{V Vashishth, B B Karak \& L Kitchatinov}            

   \maketitle

%
%

\section{Introduction}
It is believed that the convective flow of the ionized plasma in the CZs of the Sun and Sun-like stars are responsible for the generation of the magnetic field and cycles through the hydromagnetic dynamo \citep{Mof78,Cha20}. 
\blue{
In this dynamo, differential rotation and helical convective flow play important roles.
}
This is because the differential rotation of the star generates the toroidal field from the poloidal one through the so-called $\omega$ effect. On the other hand, the helical turbulence induces the poloidal field from the toroidal one which is popularly known as the $\alpha$ effect. 
In this type of  $\alpha$-$\omega$ dynamo model, the governing parameter is the dynamo number, which is defined as,
\begin{equation}
{\bf D} = \frac{\alpha \Delta \Omega R^3}{\eta^2},
\end{equation}
where, $\alpha$  is the measure of $\alpha$-effect, $\Delta\Omega$ is the variation in the 
angular velocity in the sun/star, R is its radius, and $\eta$ is the turbulent magnetic diffusivity \citep{KR80}. There is a critical dynamo number ($D_c$) below which the dynamo is not possible and the initial magnetic field decay. The regime below $D_c$ is known as the subcritical regime and above $D_c$ is called the supercritical regime \citep{Choudhuri_Book, KKV21}.

Since the rotation rate of a star decreases with the age \citep{Skumanich72, R84}, the dynamo number $D$ is expected to decrease as the star spins down \citep{KN17}. Therefore, the question is, will the dynamo cease immediately when $D < D_c$?
Interestingly, it has been found that the dynamo is still possible when $D < D_c$. \citet{KO10} have shown this subcritical dynamo in a kinematic mean-field dynamo model with non-linear quenching in $\alpha$ and $\eta$. They found that in the subcritical regime, if the dynamo is started with a strong magnetic field, a strong oscillating solution is possible. In contrast, when the dynamo is initiated with a weak field, a decaying solution is produced.  
Thus this dependence of the magnetic field with the dynamo number shows a hysteresis behaviour. 
Further, in a simplified model, \citet{KN15} showed that
transitions between
two modes (subcritical dynamo with finite magnetic field and supercritical dynamo) qualitatively reproduce two distinct modes in the distribution of  solar activity as inferred from cosmogenic isotope content in natural archives \citep{Uea14}.
This behaviour was further supported by \citet{KKB15} in the turbulent dynamo simulations.

In the above-mentioned study \citep{KO10}, a helical $\alpha$, distributed over the whole CZ was used. However, recently, there are observational supports for the predominance of the \bl\ process for the generation of a poloidal field in the Sun \citep{Das10, KO11, Muno13, Priy14, CS15}. Various surface flux transport \citep{wang89, Bau04, Ji14} and dynamo models \citep{Kar14a, C18, Cha20} based on this \bl\ process alone have been successful in modeling various aspects of solar magnetic fields and cycles.  
Therefore, in this study, we shall explore the subcritical dynamo in a \bl\ type solar dynamo model. 

As our model is kinematic, we do not capture any nonlinearity in the mean flows. We however consider magnetic field 
dependence nonlinearity in turbulent diffusivity and \bl\ $\alpha$. While diffusivity quenching is obvious and 
we have some estimates based on certain approximations \citep{Kit94, Kar14b}, the quenching in \bl\ $\alpha$ 
is less constrained. In \bl\ process, a poloidal field is produced by the decay and the dispersal of tilted bipolar magnetic regions (BMRs). This tilt has some magnetic field dependence, although its exact dependence is not well constrained \citep{Das10, Jha20}. There is also a latitudinal variation of BMRs with the solar cycle \citep{MKB17} which may be a source of nonlinear quenching \citep{J20, Kar20}. In our study, we shall consider 
magnetic field dependent quenching in both diffusivity 
and $\alpha$ based on quasi-linear approximation as presented in \citet{RK93, Kit94}. 

In the \bl\ $\alpha$, there are some inherent randomness as primarily seen in the tilts of BMRs around Joy's law \citep{Das10, SK12, MNL14, Wang15, Arlt16, Jha20}. These fluctuations can have a serious impact on the magnetic cycle and particularly on the existence of the subcritical dynamo branch.
Therefore we shall also include the fluctuations in the \bl\ $\alpha$ term of our dynamo model and check the dynamo behaviour in different regimes.

\section{Model}
For our study, we consider the magnetic field to be axisymmetric and thus we express it in the following form
\begin{equation}
{\bf B_{\rm total}} = {\bf B_p} + {\bf B_\phi} =\nabla \times [ A(r, \theta, t) \phihat] + B (r, \theta, t) \phihat,
\end{equation}
where ${\bf B_p} = \nabla \times [ A \phihat]$ 
is the poloidal component of the magnetic field and
$B$ is the toroidal component.
The evolutions of the poloidal and toroidal fields  take the following forms.
\begin{equation}
\frac{\partial A}{\partial t} + \frac{1}{s}({\bf v_p}.\nabla)(s A)
= \eta_T \left( \nabla^2 - \frac{1}{s^2} \right) A + S(r, \theta; B),
\label{eqpol}
\end{equation}

\begin{equation}
\frac{\partial B}{\partial t}
+ \frac{1}{r} \left[ \frac{\partial}{\partial r}
(r v_r B) + \frac{\partial}{\partial \theta}(v_{\theta} B) \right]
= \eta_T \left( \nabla^2 - \frac{1}{s^2} \right) B + s({\bf B_p}.{\bf \nabla})\Omega + \frac{1}{r}\frac{d\eta}{dr}\frac{\partial{(rB)}}{\partial{r}},
\label{eqtor}
\end{equation}

where $s = r \sin \theta$, ${\bf v_p} = v_r {\bf \hat{ r}} + v_\theta {\bf \hat{ \theta}}$ is the meridional flow, which is obtained through observationally-guided analytic formula as given in \citet{KC16}, $S$ is the source for the 
poloidal field. 
In the classical $\alpha \Omega$ mean-field model, $S$ is due to the helical nature of the convective flow. However, in the case of \bl\ process, the poloidal field is generated near the surface through the decay and dispersal of tilted BMRs.
In our axisymmetric model, this process has been routinely parameterised as
\begin{equation}
 S(r, \theta; B) = \alpha_{\rm BL} \mean B (\theta,t),
\label{BLsource}
\end{equation}
where $\mean B(\theta,t)$ is the average  toroidal field in a thin layer at the base of the CZ (BCZ) ($0.675\Rs < r < 0.725\Rs$) and 
$\alpha_{\rm BL}$ is the parameter for Babcock--Leighton process. 
We write,
\blue{
\begin{eqnarray}
\alpha_{\rm BL} =  \alpha \phi_\alpha( \beta ),
\label{eq:alqu}
\end{eqnarray} 
where $\beta = B/B_0$ with $B_0$ being the equipartition field strength.
}
The $\alpha$ is the profile for the usual \bl\ $\alpha$
which has the following form:
\begin{eqnarray}
    \alpha = \frac{\alpha_0}{4} \left[ 1+\erf \left(\frac{r - 0.95R_\odot}{0.05 R_\odot}\right)\right] \times \left[1-\erf \left(\frac{r - R_\odot}{0.01 R_\odot}\right)\right] 
    \frac{\sin\theta\cos\theta}{1+\exp(K)},
\label{alphaprof}
\end{eqnarray}
for $\theta < \pi/2$, $K = 30(\pi/4-\theta)$ and for $\theta > \pi/2$, $K= 30(\theta-3\pi/4)$.

The $\phi_\alpha(\beta)$ in \Eq{eq:alqu} has the following magnetic field dependent quenching
based on  quasi-linear approximation as presented in \citet{RK93}.   
\blue{
\begin{equation}
{\phi_\alpha({\beta})}= \frac{15}{32{\beta}^4} \left[ 1-\frac{4{\beta}^2}{3(1+{\beta}^2)^2}-
\frac{1-{\beta}^2}{{\beta}}\arctan {\beta} \right].
\label{eq:phi_alpha}
\end{equation}
}
The turbulent diffusivity $\eta_T$ has the following form.
\begin{equation}
\eta_T = \eta \phi_\eta({\beta}), 
\label{eq:etaqu}
\end{equation}
where
\begin{equation}
\eta = \eta_0 \left[ \eta_{in}+ \left(\frac{1-\eta_{in}}{2}\right)
\left( 1 +\erf\left(\frac{r-x_n}{h_n}\right)\right)\right]
\end{equation}
with $\eta_{in}=10^{-4}$, $x_n=0.70 R_\odot$, $h_n=0.05 R_\odot$, $\eta_0= 5 \times 10^{12}$~\cmss.
We consider a magnetic field dependent quenching in the diffusivity following
\citet{KPR94}
\blue{
\begin{equation}
{\phi_\eta({\beta})}=\frac{3}{8{\beta}^2}\left[ 1 + \frac{4+8{\beta}^2}{(1+{\beta}^2)^2}+
\frac{{\beta}^2-5}{{\beta}}\arctan {\beta} \right].
\label{eq:phi_eta}
\end{equation}
}
We note that both $\alpha$ and $\eta_T$ are quenched through the local magnetic field.
In \Sec{sec:nonlocalquench}, however, we shall change this prescription and relate $\alpha$ and $\eta_T$
through the magnetic field at the BCZ.

The differential rotation $\Omega$ and the boundary conditions
are the same as that used in \citet{NC02, CNC04}.


\section{Results}
\subsection{Quenching with the local field}
\begin{figure}
    \centering
    \includegraphics[width=1\columnwidth]{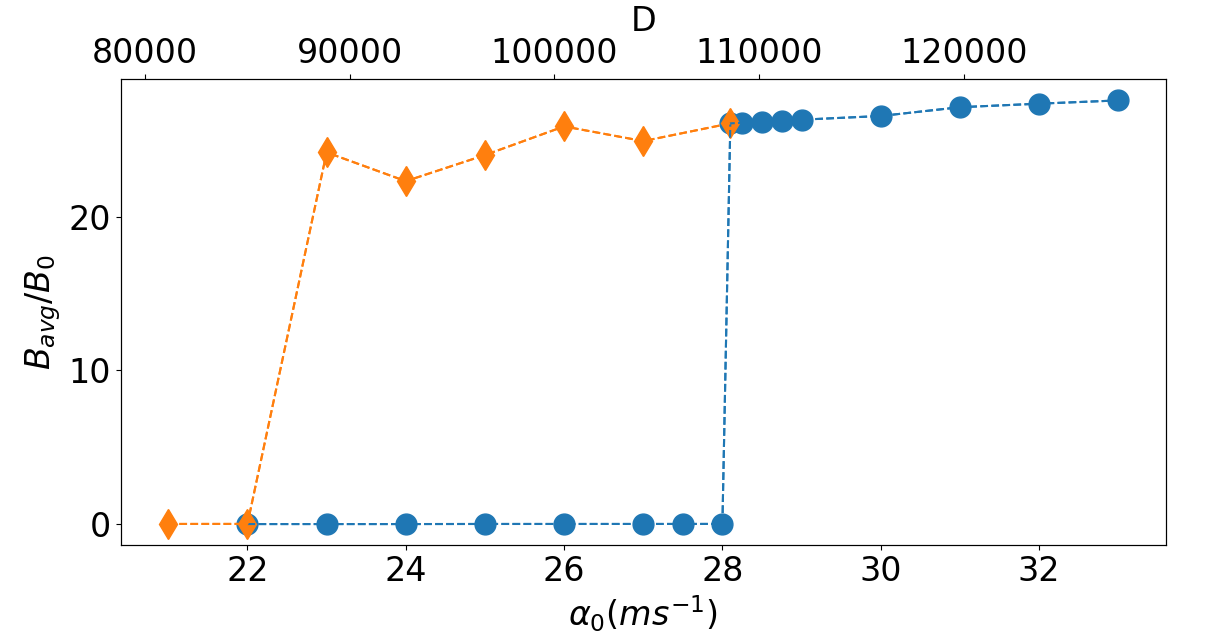}
    \caption{Dynamo hysteresis: Variation of the temporal average of the mean toroidal field normalized to $B_0$ and computed 
    at the BCZ at latitude $-14^\circ$ ($B_{avg}$) 
    as a function of $\alpha_0$ (in \mps) from simulations started with a weak field (filled circles) and from simulations started with strong field of previous simulation 
    (orange diamonds).  The corresponding $D$ is shown in top horizontal axis.
        }
    \label{fig:hyst}
\end{figure}

By including the magnetic field dependent quenching in $\eta_T$ and $\alpha$ 
and by specifying the large-scale flows, such as differential rotation and meridional circulation, we solve the dynamo \Eqs{eqpol}{eqtor}.
We first identify the dynamo transition. To do so, we perform simulations at different values 
of $\alpha_0$, i.e., at
different values of dynamo number, defined here as $D = \alpha_0 \Omega_0 R^3 /\eta_0^2$ 
(where $\Omega_0 = 2 \pi /T$; $T=25.38$~days).
We find that when we start the simulations with a very weak magnetic field, 
the initial magnetic field grows as long as $\alpha_0 \ge 28.10$~\mps.
The dynamo number corresponding to this $\alpha_0$, i.e., the critical dynamo number, $D_c = 1.086\times 10^5$.
The points connecting the solid line in \Fig{fig:hyst}
show $B_{\rm avg}$
versus $\alpha_0$ and $D$. $B_{\rm avg}$ is computed at $0.7\Rs$ and $-13^\circ$ latitude and averaged over a few steady cycles.
Now, instead of starting the simulation with a weak magnetic field, we start 
it with the output of an oscillatory solution of a strong magnetic field. 
We take the output of the simulation
performed at $\alpha_0=28.10$~\mps and 
execute a new simulation 
at $\alpha_0=27$~\mps
and then we take the output of this simulation and feed it into a new simulation at a lower $\alpha_0$.
In this way, we perform several simulations at a progressively lower $\alpha_0$ by taking the output of the previous simulation at higher $\alpha_0$.
The orange diamond points connecting the dotted line in \Fig{fig:hyst}
shows $B_{avg}$ for these simulations.
The interesting behaviour we observe is that the solutions are different than the ones performed at the same parameters but 
started with a weak field. 
We find a wide region in the dynamo parameter space, as shown in \Fig{fig:hyst}, over which
the dynamo is decaying when started with the weak field 
but produces a strong oscillatory field when started with
a strong field. Overall the dynamo shows a hysteresis behaviour. 
This behaviour and the subcritical dynamo, for the first time,
was discovered in a simple mean-field dynamo with distributed $\alpha$ \citep{KO10}. 
Later dynamo hysteresis was confirmed in magnetohydrodynamics (MHD) simulations of the helical turbulent dynamo with imposed shear \citep{KKB15}. Recently this was also seen in numerical simulations of turbulent $\alpha^2$ dynamo \citep{oliveira20}.

As discussed in \citet{KO10}, the imposed magnetic-field dependent nonlinearity in turbulent diffusivity and $\alpha$
makes the behaviour of the effective dynamo number
$D_{\rm eff} = \alpha_0 \phi_\alpha (\beta) \Omega R^3 /\eta_0 ^2 \phi_\eta(\beta) ^2$ non-monotonic.
When the magnetic field is large ($\beta \gg 1$), $D_{\rm eff}$ decreases with the increase of $\beta$ ($D_{\rm eff} \sim \beta^{-1}$). In contrast, 
when $\beta$ is small  ($\beta \ll 1$), $D_{\rm eff} $ increases with the increase of $\beta$ ($D_{\rm eff} \sim D(1 + 16 \beta^2/7)$); see Fig.\ 1 of \citet{KO10}.
Thus, when the simulation is started with a very weak field, $D_{\rm eff}$ remains small and cannot trigger the dynamo. On the other hand, when the simulation is started with a strong field ($\beta\sim 1$), $D_{\rm eff}$ becomes large enough to
produce dynamo action.

\begin{figure}
    \centering
    \includegraphics[width=1.05\columnwidth]{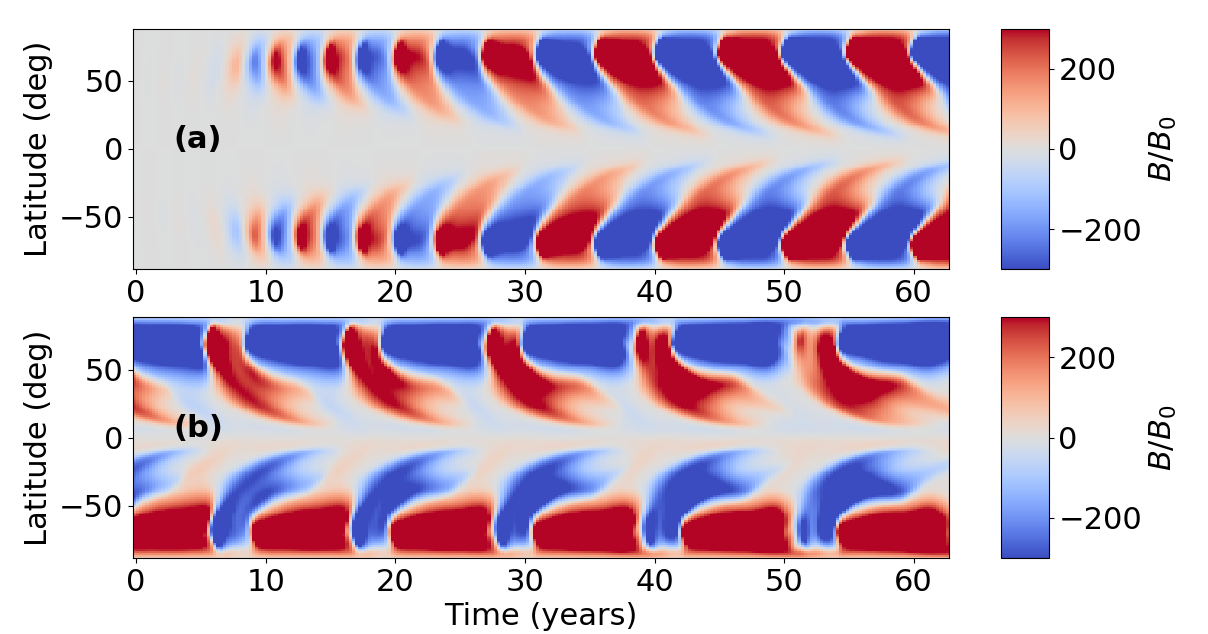}
    
    \caption{
    (a) Time-latitude distribution of the toroidal field at BCZ from a dynamo simulation at critical dynamo case ($\alpha = 28.10~$\mps) for which the simulation started with a weak field.
    (b) Same as (a) but from a subcritical dynamo case ($\alpha = 23~$\mps) and the simulation
    started with a strong field.
    }
    \label{fig:bfly}
\end{figure}

The time-latitude distributions of the toroidal magnetic fields from a simulation at 
critical $\alpha_0 = 28.10$~\mps\ (started with a weak field) 
and from a subcritical case at $\alpha_0 = 23$~\mps (started with a strong initial field) 
are shown in \Fig{fig:bfly}.
We observe regular polarity reversal and some migration towards the equator. The cycle period is much shorter than the solar value. 
This short cycle is due to our chosen value of $\eta_0$ ( = $5\times 10^{12}$~\cmss). The high diffusivity always tends to produce short cycle \citep{KC12, KC16} unless we reduce the diffusivity at BCZ drastically and/or include a strong downward 
magnetic pumping \citep{KO12, KC16}. In fact, if we do not include
the nonlinearity in diffusivity, then the cycle period is 
even shorter ($\sim 0.83$ years).

One aspect of all these simulations is that they produce unexpectedly 
strong magnetic field near the BCZ. In \Figs{fig:hyst}{fig:bfly} we observe that the 
magnetic field strength is several tens of stronger than $B_0$. 
This strong field is caused by the strongly quenched diffusivity near the 
BCZ. We recall from \Eqs{eq:alqu}{eq:etaqu} that $\alpha_{\rm BL}$
and $\eta_{\rm T}$ are related to magnetic field locally.
Near the BCZ, the magnetic field is usually stronger than that near the surface and thus at the BCZ, $\eta_{\rm T}$
is reduced strongly, but $\alpha_{\rm BL}$ is zero there. 
Hence this strongly reduced diffusivity near the BCZ in our
\bl\ type dynamo is causing this strong magnetic field.
This strong magnetic field is indeed in agreement with the super-equipartition field which was a prediction of the thin flux-tube simulations \citep{DC93, CMS95}.

\subsection{Quenching with the non-local field}
\label{sec:nonlocalquench}
\subsubsection{Regular dynamo solutions and hysteresis}
The \bl\ $\alpha$ is a nonlocal process in which the magnetic field
at the BCZ acts as the seed for the poloidal field; see \Eq{BLsource}. Therefore, instead of connecting $\alpha_{\rm BL}$
and $\eta_{\rm T}$ with the local magnetic field, we now connect them with the magnetic field at the BCZ, i.e.,
\begin{eqnarray}
\alpha_{\rm BL} =  \alpha \phi_\alpha(\mean \beta )
\label{eq:alphaquavg}
\end{eqnarray}
\begin{eqnarray}
\eta_T = \eta \phi_\eta(\mean{\beta}), 
\label{eq:etaquavg}
\end{eqnarray}
where $\mean\beta = \mean B/B_0$.
The quenching functions $\phi_\alpha$ and $\phi_\eta$ will be computed from the same \Eqs{eq:phi_alpha}{eq:phi_eta} but based on the average toroidal field at BCZ ($\mean B$ ).
No other changes are made in the model.

\begin{figure}
    \centering
    \includegraphics[width=1\columnwidth]{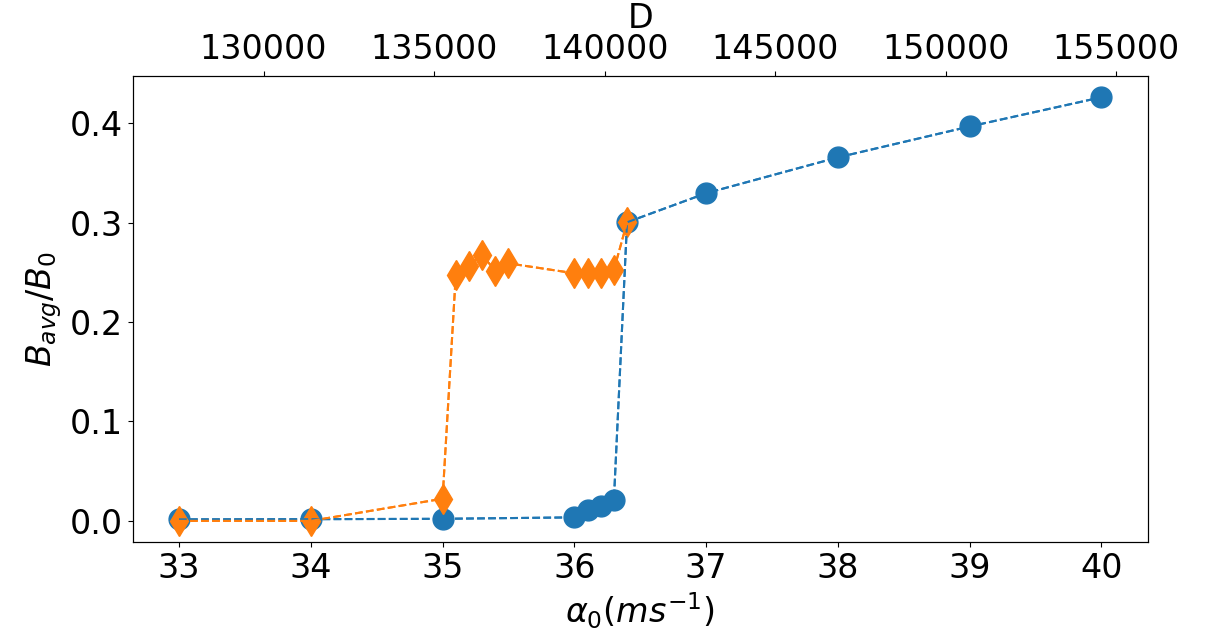}
    \caption{Same as \Fig{fig:hyst} but obtained from 
    simulations in which $\alpha_{\rm BL}$ and $\eta_T$
    are related to the toroidal field at the BCZ.
        }
    \label{fig:hyst_avg}
\end{figure}

\begin{figure}
    \centering
    \includegraphics[width=1.05\columnwidth]{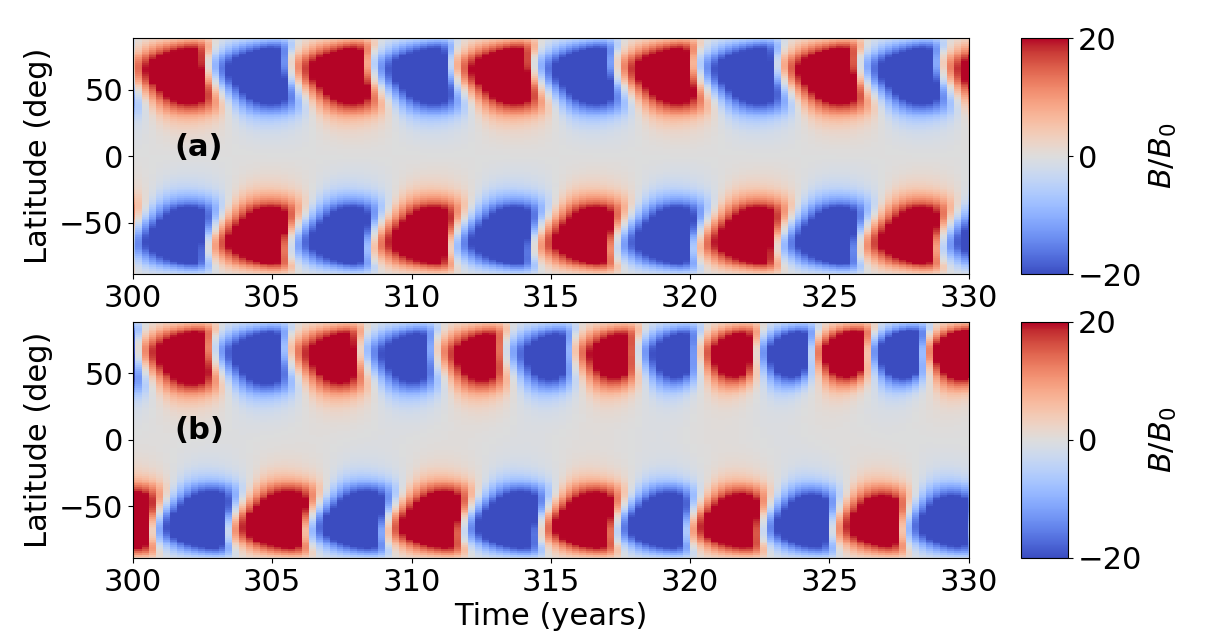}
    \caption{Butterfly diagrams for (a) toroidal field at BCZ for critical dynamo for the case ($\alpha_0 = 36.40$~\mps) when the simulation started with a weak magnetic field and (b) for subcritical dynamo case ($\alpha_0 = 35$~\mps) and started with the output of a strong oscillatory solution at $\alpha_0 = 36$~\mps. 
 }
    \label{fig:critbflyweak}
\end{figure}

We perform the simulations at different values of $\alpha_0$ in the same way as we have done 
to produce \Fig{fig:hyst}. \Fig{fig:hyst_avg}, shows the results. 
We immediately notice that the magnetic field strength 
is reduced, at least by an order of magnitude.
We again find a regime
in the dynamo parameter where two solutions are possible: 
a weak decaying field and a strong oscillatory field, 
depending on the initial condition.  Thus, 
the dynamo hysteresis is a generic feature in the \bl\ type solar dynamo.

\Fig{fig:critbflyweak} shows the time-latitude distribution of the 
 toroidal field from simulation at the critical $\alpha_0 = 36.40$~\mps (started with a weak field) and at the subcritical dynamo, $\alpha_0 = 35$~\mps (started with a strong  oscillating field). 
 Although we see some general features of the solar magnetic field in this simulation, the cycle period is considerably reduced. The average period of the magnetic field oscillation is about 2.5 years.
This short cycle period is due to different nonlinear quenching in $\alpha$ and $\eta$. Further, the field is strongest near the poles.

\subsubsection{Dynamo with fluctuations in $\alpha$}
So far in each simulation, all dynamo parameters were kept constant and thus the nonlinearity in our model kept the amplitude of the magnetic cycle nearly equal.  However, due to fluctuating nature of the stellar convection, the dynamo parameter, especially the $\alpha$ is subjected to fluctuate around its mean. In the Babcock--Leighton scenario, the fluctuations are primarily seen in the form of scatter in the bipolar active region tilts around Joy's law \citep[e.g.,][]{SK12, MNL14, Wang15, Arlt16, Jha20} and the randomness in flux emergence \citep{KM17}. On the other hand, in the turbulent mean-field $\alpha$, the scatter is unavoidable due to finite numbers of convection cells \citep{C92}.   
The fluctuations in $\alpha$ cause the polar field to change and thus make the magnetic cycle unequal as
observed in sun and sun-like stars.
This has been already used in many studies
for modeling the irregular aspects of solar cycles \citep[e.g.,][]{CD00, CCJ07, CK09, KC11, OK13, KMB18}.

Motivated by this, we include fluctuations in our \bl\ $\alpha$. To do so, we replace $\alpha_0$ by $\alpha_0 = \alpha_0 [ 1 + s (\tau_{\rm corr}) \times f]$, where $s$ is the uniform random number in the interval $-1 < s < 1$ and $\tau_{\rm corr}$ is the 
coherence time, which is taken as one month---consistent with the mean lifetime of BMRs. Thus, now in our model, the value of $\alpha_0$ is updated randomly every one month. The level of fluctuations is determined by $f$. For example, $f=1$, and 0.2 correspond to $100\%$ and $20\%$ fluctuations, respectively.

\begin{figure*}
    \centering
    \includegraphics[width=1\columnwidth]{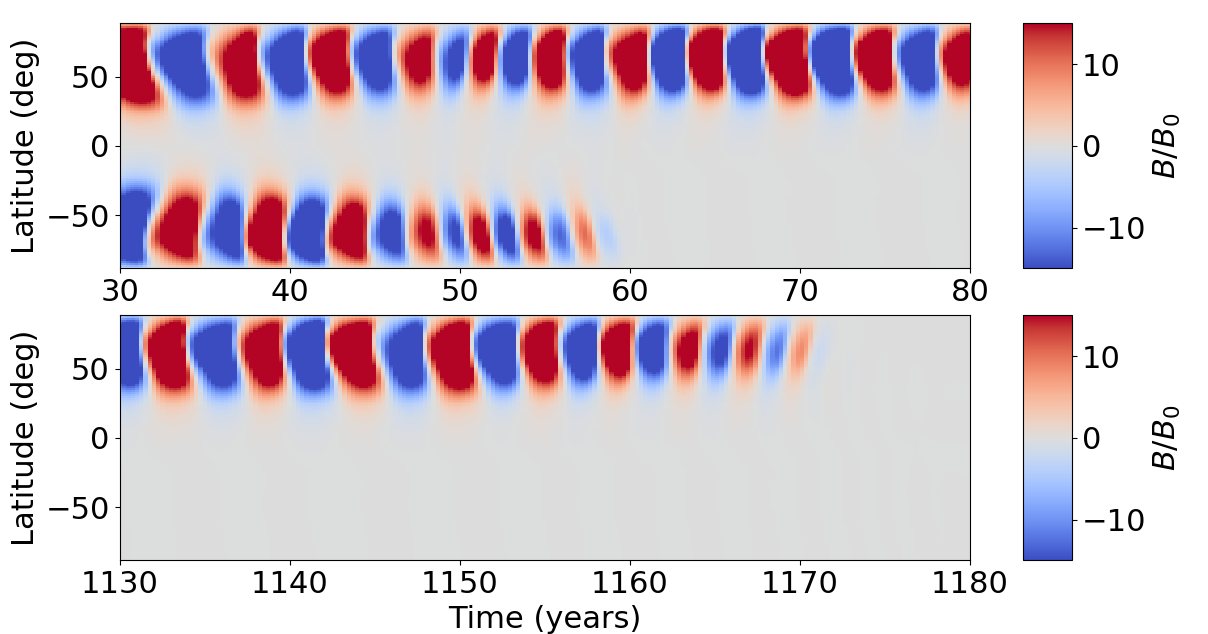}
    \caption{Butterfly diagrams of  toroidal field for subcritical dynamo at $\alpha_0 = 35$~\mps with 20 percent fluctuations. Note that the duration 80--1130 years, i.e., the time spans shown in between two panels are not displayed.
    }
    \label{fig:bflysubcritFl}
\end{figure*}

We find that for subcritical and slightly above critical regimes, this model tends to decay at  large fluctuations. The dynamo dies even at $10\%$ fluctuations. \Fig{fig:bflysubcritFl} shows the butterfly diagram of a subcritical dynamo ($\alpha_0 = 35$~\mps) in which
the magnetic field decayed after about 1000 years due to large fluctuations. We note that this did not happen simultaneously in two hemispheres. Thus the subcritical branch is unstable under the large fluctuations. 
We have checked that if the fluctuation level is below 10$\%$, then the dynamo does not decay immediately; sometimes it decays in a few years and sometimes it produces cycles for thousands of years before the decay. This is surprising. However, this problem might be solved by adding a distributed $\alpha_0$ in the 
CZ which has been a way for recovering the dynamo from a grand minimum \citep{KC13, Ha14}.

However, in the supercritical regime, the dynamo maintains a stable solution even at a very large fluctuations.
We observe that when we have included 20$\%$ fluctuations, subcritical and critical cases die whereas supercritical case $\alpha_0 = 40$~\mps survives. Hence, the critical dynamo number increases with the increase of the level of fluctuations.

\begin{figure}
    \centering
    \includegraphics[width=1\columnwidth, height=8cm]{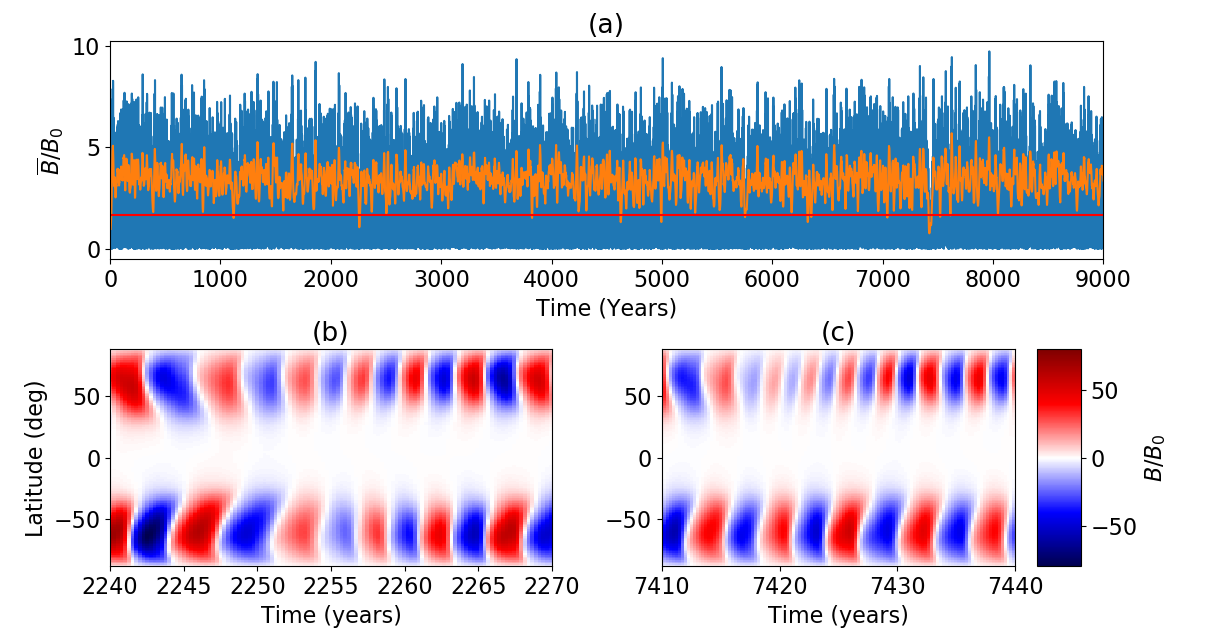}
    \caption{(a) Time series plot along with its smoothed curve (yellow) of toroidal magnetic flux. Red horizontal line shows the half of the mean of this smooth curve. (b) $\&$ (c) Butterfly diagram of toroidal field for the two grand minima. These are obtained from a supercritical dynamo at $\alpha_0 = 40$~\mps}
    \label{fig:TSsupercritical}
\end{figure}

The time series of the toroidal magnetic flux $\overline{B}$ at the BCZ from a simulation for 9000 years at $\alpha_0 = 40$~\mps\ is shown in \Fig{fig:TSsupercritical}(a). $\overline{B}$ is computed in a small region with $r = 0.677\Rs$--0.726$\Rs$ and latitudes: 10$^\circ$--45$^\circ$. We can see that the cycles are now variable, occasionally producing significantly strong and weak cycles. To check whether this simulation produces any 
grand minima or not, we apply the same method as performed in \citet{USK07} for the Sun. We bin the data for the duration of one cycle period (which is about 2.5 years in this simulation),
filter the data by using Gleissberg's low-pass filter 1-2-2-2-1,
and finally, count a grand minimum if this smoothed data falls below 
50$\%$ of its mean for at least two cycle periods, which is 5 years
in our case. In this way, we detect 6 grand minima.
Two of such cases are presented in \Fig{fig:TSsupercritical}(b) and (c). 
When we increase the supercriticality of the model by increasing $\alpha_0$, the number of grand minima decreases. When $\alpha_0 \ge$ 42 \mps\ we do not observe any grand minima. 
This is in someway agreement with the stellar observations 
\blue{
because the only slowly rotating stars produce grand minima \citep{Baliu95} and the slowly rotating stars are expected to have smaller 
value of $\alpha_0$. This is due to the fact that the efficiency of the 
\bl\ process depends on the tilt which is rooted 
to the rotation of the star \citep{DC93}.
}

\begin{figure}
    \centering
    \includegraphics[width=8cm, height=5cm]{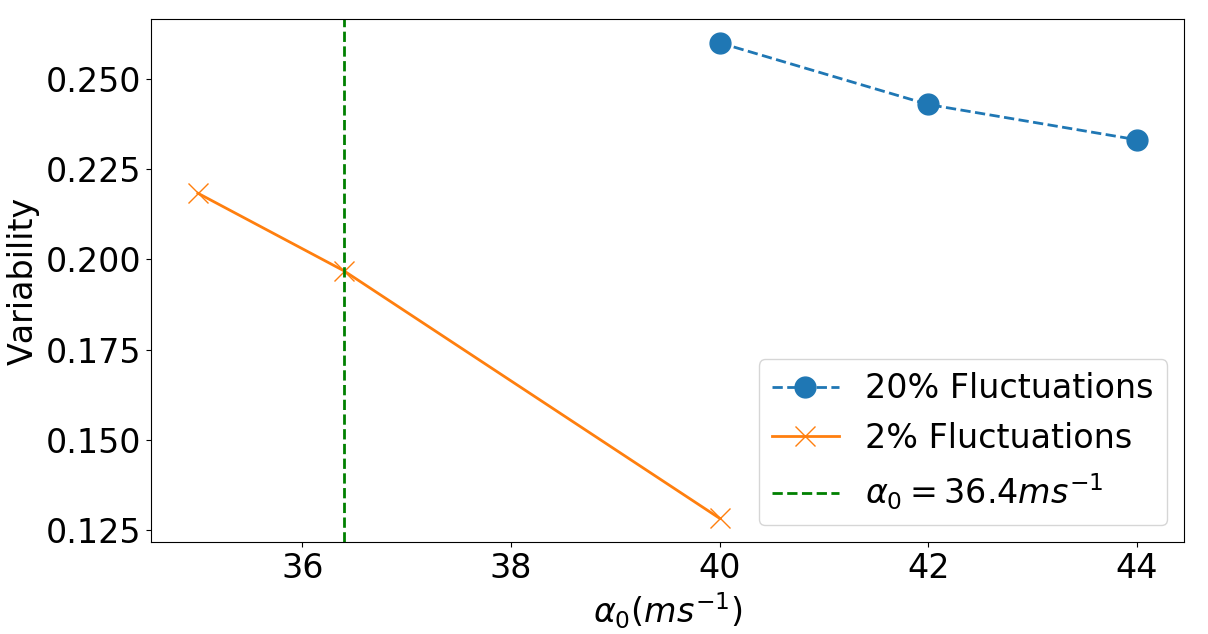}
    \caption{Variation of variability with respect to the increase in $\alpha_0$  with $2\%$  fluctuations (orange solid line) and with $20\%$ fluctuations (blue dashed line).
    }
    \label{fig:variability}
\end{figure}

The amount of variability of the cycle is obviously more when the fluctuation is more; see \Fig{fig:variability}. 
To compute the variability, we first compute the peaks of the cycles as measured from the toroidal magnetic field time series {$\overline{B}$}. Then the root mean-square of the peaks divided by the mean is taken as the variability.
The variability decreases with the increase of supercriticality of the model ($\alpha_0$).

\section{Conclusions}

In this work, we have applied an axisymmetric kinematic solar dynamo model with 
a \bl\ $\alpha_{\rm BL}$ as the source of the poloidal field to explore the 
subcritical dynamo and the hysteresis behaviour. We have included magnetic field
dependent nonlinearities in the $\alpha_{\rm BL}$ and diffusivity $\eta_T$ based on the quasi-linear approximation \citep{RK93, Kit94}. 
We have included these nonlinearities in two ways. First,
we connect the $\alpha_{\rm BL}$ and $\eta_T$ with the local toroidal magnetic field 
and in the second,
we connect these with the toroidal field at the BCZ.
We find regular polarity reversals and cycles as long as the dynamo number is above a critical value. 
We find a regime in the dynamo parameter where two solutions are possible: a weak decaying field and a strong oscillatory field, depending on the initial condition. Hence, 
the dynamo hysteresis, which was predicted in the distributed $\alpha$ $\Omega$ dynamo \citep{KO10} and turbulent dynamo simulations \citep{KKB15}, 
also survives in \bl\ type dynamos.
Thus, our study along with previous studies, provide a possible
existence of subcritical dynamo for the execution of large-scale magnetic cycles in sun-like stars.

By including stochastic fluctuations in $\alpha_{\rm BL}$, we check the stability of these subcritical branches. We find that
 when $\alpha_{\rm BL}$ and $\eta_T$ are connected with the local magnetic field,
 the subcritical branch maintains stable magnetic cycles. However, 
 in the other case, when the $\alpha_{\rm BL}$ and $\eta_T$ are connected with
 the magnetic field at the BCZ, the subcritical branch tends to decay with fluctuations.
The supercritical branch is always stable and produces some grand minima. The number of grand minima and the variability of the cycle decrease with the increase of supercriticality (as controlled by the strength of $\alpha_{\rm BL}$ in our case). 

\begin{acknowledgements}
The authors thank Pawan Kumar and the referee for carefully reviewing the manuscript.
Financial Support from the Department of Science and Technology (SERB/DST), India through the Ramanujan fellowship (project No. SB/S2/RJN-017/2018) awarded to BBK is acknowledged. B.B.K. also acknowledges
the funding provided by the Alexander von Humboldt
Foundation.VV acknowledges the financial support from the DST through INSPIRE fellowship. LK is thankful for the support from the Russian Foundation for Basic Research (Project 19-52-45002\_lnd) and the ministry of Science $\&$ High Education of the Russian federation. 
\end{acknowledgements}

\bibliographystyle{raa}
\bibliography{paper}

\end{document}

%% file: newcommands.tex
\newcommand{\EQ}{\begin{equation}}
\newcommand{\EN}{\end{equation}}
\newcommand{\EQA}{\begin{eqnarray}}
\newcommand{\ENA}{\end{eqnarray}}

\newcommand{\mps}{m~s$^{-1}$}

\newcommand{\cmss}{cm$^2$~s$^{-1}$}

\def\bl{Babcock--Leighton}

\def\Rs{R_{\odot}}

\newcommand{\phihat}{\hat{\mbox{\boldmath $\phi$}} {}}

\newcommand{\mean}[1]{\overline{#1}}

\newcommand{\Eq}[1]{Equation~(\ref{#1})}
\newcommand{\Eqs}[2]{equations~(\ref{#1}) and~(\ref{#2})}

\newcommand{\Sec}[1]{ \S \ref{#1}}

\newcommand{\Fig}[1]{Figure~\ref{#1}}

\newcommand{\Figs}[2]{Figures~\ref{#1} and \ref{#2}}